\documentclass[11pt,a4paper]{article}
\usepackage{jheppub,amsmath,amssymb,slashed,url,bm}
\usepackage{graphicx}
\usepackage{epstopdf}

\newcommand{\e}{\begin{eqnarray}}
\newcommand{\ee}{\end{eqnarray}}
\newcommand{\CN}{{\cal N}}

\def\d{\delta}
\newcommand{\ep}{\epsilon}

\newcommand{\vp}{\varepsilon}

\font\teneurm=eurm10 \font\seveneurm=eurm7  \font\fiveeurm=eurm5
\newfam\eurmfam
\textfont\eurmfam=\teneurm \scriptfont\eurmfam=\seveneurm
\scriptscriptfont\eurmfam=\fiveeurm

\font\teneusm=eusm10 \font\seveneusm=eusm7 \font\fiveeusm=eusm5
\newfam\eusmfam
\textfont\eusmfam=\teneusm \scriptfont\eusmfam=\seveneusm
\scriptscriptfont\eusmfam=\fiveeusm

\font\tencmmib=cmmib10 \skewchar\tencmmib='177
\font\sevencmmib=cmmib7 \skewchar\sevencmmib='177
\font\fivecmmib=cmmib5 \skewchar\fivecmmib='177
\newfam\cmmibfam
\textfont\cmmibfam=\tencmmib \scriptfont\cmmibfam=\sevencmmib
\scriptscriptfont\cmmibfam=\fivecmmib

\title{Constructing Nonabelian $(1,0)$ Hypermultiplet Theories in Six Dimensions}

 \author{Fa-Min Chen }
\affiliation{Department of Physics, Beijing Jiaotong University, Beijing 100044, China}
\abstract{We construct a class of nonabelian superconformal $(1,0)$ hypermultiplet
theories in six dimensions by introducing an abelian auxiliary field. The gauge fields
of this class of theories are non-dynamical, and this class of theories can be viewed as
 Chern-Simons-matter theories in 6D.}% We also briefly discuss the application of the theories to little strings.}
\begin{document} \maketitle

\section{Introduction}\label{secintro}
It is known that six is the maximal dimension for a superconformal field theory, and the superalgebra of the superconformal symmetry in 6D is $OSp(8|2)$ or $OSp(8|4)$  \cite{Nahm}. 6D superconformal theories are interested because they can be the candidates of dual gauge theories of multiple D5 or M5-branes \cite{Juan}, or may be used to study the dynamics of compactifications of superstring theory to 6D \cite{Seiberg}.

Recently, we have constructed a class of nonabelian $(1,0)$ tensor multiplet theories
\cite{chen17}, which has the same field content as that of the nonabelian $(2,0)$ tensor multiplet theory of Lambert and Papageorgakis (LP) \cite{LP}. Classically, these $(1,0)$ tensor multiplet theories have an $OSp(8|2)$ superconformal symmetry.

In this paper we work on another direction: we construct a class of nonabelian
superconformal $(1,0)$ hypermultiplet theories in 6 dimensions; the gauge fields
of the theories are \emph{nondynamical}.
\footnote{6D field theories of hypermultiplets coupled to \emph{dynamical} vector and tensor mutiplets have been constructed in \cite{Sezgin3}; 6D field theories of hypermultiplets coupled to Yang-Mills fields involving higher derivatives have been constructed in \cite{Smilga1} and investigated in \cite{Simlga2} and \cite{Simlga3}.}.
They can be thought of as a class of Chern-Simons-matter theories in 6D. Classically, this class of theories also have an $OSp(8|2)$ superconformal symmetry. Nevertheless, they should be simpler than the nonabelian $(1,0)$ tensor multiplet theories in Ref. \cite{chen17}, since they do not involve the self-dual strengths $H_{\mu\nu\rho}$ associated with the tensor multiplets. On the other hand, it seems that this class of theories cannot be derived by truncating any known nonabelian $(1,0)$ or $(2,0)$ tensor multiplet theories. So it may be interesting to study this class of $(1,0)$ hypermultiplet theories. In particular, it would be interesting to quantize this class of theories and to construct their gravity duals. We shall present the details of construction of this class of theories in the next section.

\section{Nonabelian Superconformal $(1,0)$ Theories of Hypermultiplets}
%\subsection{Closure of the Superalgebra}
We begin by reviewing the free $(1,0)$ theory of hypermultiplet in 6D. The component fields are given by  $(X^i, \psi_{-})$  (we follow the convention in Ref. \cite{chen10}). Here $X^i$ ($i=6,7,8,9$) are four real scalar fields; and the ferminoic field $\psi_-$ is defined as $\psi_-=\frac{1}{2}(1-\Gamma_{6789})\psi$, where $\psi$ is antichiral with respect to $\Gamma_{012345}$, i.e. $\Gamma_{012345}\psi=-\psi$ (we work with 32-component Majorana fermions). The scaling dimensions of the fields are the following: $[X^i]=2$ and $[\psi_-]=5/2$. The law of $(1,0)$ supersymmetry transformations is given by
\e
\d X^i&=&i\bar\ep_+\Gamma^i\psi_-,\nonumber\\
\d\psi_-&=&\Gamma^\mu\Gamma^i\ep_+\partial_\mu X^i,
\ee
where $\ep_+=\frac{1}{2}(1+\Gamma_{6789})\ep$, and $\Gamma_{012345}\ep=+\ep$. The superalgebra is closed provided that the equations of motion
\e\label{eomff}
\Gamma^\mu\partial_\mu\psi_-=0
\ee
is obeyed.

It is instructive to count the physical degrees of freedom of $\psi_-$. Generally
speaking, the Majorana spinor field $\psi_-$ might have 32 real components.
But the conditions $\Gamma_{6789}\psi_-=-\psi_-$ and $\Gamma_{012345}\psi_-=-\psi_-$ reduce the 32 components to $8$. On the other hand, taking account of the equations
of motion $\Gamma^\mu\partial_\mu\psi_-=0$, we see that $\psi_-$ has only four (real) independent components. This result is expected:  since $\psi_{-}$ and  $X^i$ ($i=6,7,8,9$) can
form a supermultiplet, they must have exactly the same physical degrees of freedom.

  %since we have exactly four real scalar fields $X^i$ ($i=6,7,8,9$).

Except for the 6D Poincare symmetry, this free $(1,0)$ theory also possesses a global symmetry $SO(4)=SU(2)_R\times SU(2)_F$, whose chirality matrix is $\Gamma_{6789}$. Here the subscripts $R$ and $F$ stand for ``R-symmetry" and ``flavor symmetry", respectively.  Specifically, denoting the set of spinor representation matrices of $SO(4)$ as $\Gamma^{ij}=\Gamma^{[i}\Gamma^{j]}$, and defining  $\Gamma^{ij}_{\pm}=\frac{1}{2}(\Gamma^{ij}\pm\frac{1}{2}\vp^{ijkl}\Gamma^{kl})$, then $\Gamma^{ij}_+$ and $\Gamma^{ij}_-$ are two sets of representation matrices of $SU(2)_F$ and $SU(2)_R$, respectively. It can be seen that $\ep_+$ transforms as a $\textbf{2}$ of $SU(2)_R$,
$\psi_-$ transforms as a $\textbf{2}$ of $SU(2)_F$, and $X^i$ transforms in the $\textbf{4}$ representation of $SO(4)$.

The above free theory can be generalized to be a class of nonabelian $(1,0)$ theories. To see this, we propose the following supersymmetry transformations
\e\label{susy}
\d X^i_m&=&i\bar\ep_+\Gamma^i\psi_{m-},\nonumber\\
\d\psi_{m-}&=&\Gamma^\mu\Gamma^i\ep_+D_\mu X^i_m+\Gamma_\lambda\Gamma^i\ep_+C^{\lambda j}[X^i,X^j]_m\nonumber\\
\d A^m_\mu&=& i\bar\ep_+\Gamma^i\psi^m_-C^i_\mu,\nonumber\\
\d C^{\mu i}&=&0.
\ee
%Here $a$ is a constant, to be determined;
Here $C^{\mu i}$ is an abelian auxiliary field\footnote{The role of $C^{\mu i}$ is similar to that of the auxiliary field $C^\mu_a$ in the LP theory \cite{LP}, where $a$ is a 3-algebra index (See also \cite{Singh1, Singh2, chen10}). The geometric meaning of $C^\mu_a$ in loop space was investigated in \cite{Saemann}. It would be interesting to find out the geometric interpretation of $C^{\mu i}$.} with scaling dimension $-1$. The scalar and  fermionic fields are in the adjoint representation of the Lie algebra of gauge symmetry; the Lie algebra can be chosen as the ADE type. The covariant derivative is defined as
\e\label{covd}
D_\mu X^i_m=\partial_\mu X^i_m+ [A_\mu, X^i]_m=\partial_\mu X^i_m+ (A_\mu)_nX^i_pf^{np}{}_m,
\ee
where $f^{np}{}_m$ are the structure constants of the Lie algebra.

Let us now examine the closure of superalgebra. We begin by considering the scalar fields. A short calculation gives
\e\label{bosonic2}
[\d_1,\d_2]X^i_m=v^\nu D_\nu X^i_m+[\Lambda, X^i]_m,
\ee
where
\e
\label{v}
v^\mu&\equiv&-2i\bar\ep_{2+}\Gamma^\mu\ep_{1+},\\
\Lambda_m&\equiv&-v^\nu C^i_\nu X^i_m.\label{lambda}
\ee
It is clear that the superalgebra is closed on $X^i_m$, up to the gauge transformation $[\Lambda, X^i]_m$.

We now consider the super-variation of spinor fields. After some algebraic steps, we obtain
\e\label{fermion2m}
[\d_1,\d_2]\psi_{m-}&=&v^\nu D_\nu\psi_{m-}+[\Lambda,\psi_{-}]_m\nonumber\\
&&-\frac{1}{2}v^\nu\Gamma_\nu(\Gamma^\mu D_\mu\psi_{m-}+\Gamma^\mu C^i_\mu[\psi_{-},X^i]_m).
\ee
To close the superalgebra, we impose the equations of motion
\e\label{EOMp3}
0&=&\Gamma^\mu D_\mu\psi_{m-}+\Gamma^\mu C^i_\mu[\psi_{-},X^i]_m.
\ee
In deriving (\ref{fermion2m}), we have used the Fierz identity \cite{chen10}
\begin{equation}\label{Fierz2}
(\bar\ep_{2+}\chi_+)\ep_{1+}-(\bar\ep_{1+}\chi_+)\ep_{2+}=-\frac{1}{4}
(\bar\ep_{2+}\Gamma_\mu\ep_{1+})\Gamma^\mu\chi_+
-\frac{1}{192}(\bar\ep_{2+}\Gamma_{\mu\nu\lambda}
\Gamma^{ij}_-\ep_{1+})\Gamma^{\mu\nu\lambda}\Gamma^{ij}_-\chi_+.
\end{equation}
Here $\chi_+$ satisfies  the conditions $\Gamma_{012345}\chi_+=-\chi_+$ and
$\Gamma_{6789}\chi_+=+\chi_+$.

As for the auxiliary field, we have
\e\label{aux}
[\d_1,\d_2]C^{\mu i}=0=v^\nu D_\nu C^{\mu i}+[\Lambda, C^{\mu i}].
\ee
The commutator $[\Lambda, C^{\mu i}]$ vanishes due to that $C^{\mu i}$ is an abelian field. As a result, we have
\e
D_\nu C^{\mu i}=\partial_\nu C^{\mu i}=0,
\ee
i.e. $C^\mu$ is a \emph{constant} abelian field.

Finally, the commutator $[\d_1,\d_2]A^m_\mu$ is given by
\e\label{gauge2}
[\d_1,\d_2]A^m_\mu&=&v^\nu F^m_{\nu\mu}-D_\mu\Lambda^m\nonumber\\
&&+v^\nu(F^m_{\mu\nu}+2C^i_{[\mu}D_{\nu]}(X^i)^m+C^i_\mu C^j_\nu[X^i,X^j]^m),
\ee
where $F^m_{\mu\nu}=\partial_\mu A^m_\nu-\partial_\nu A^m_\mu+[A_\mu,A_\nu]^m$. We see that the second line of (\ref{gauge2}) must be the equations of motion for gauge fields
\e
0&=&F^m_{\mu\nu}+2C^i_{[\mu}D_{\nu]}(X^i)^m+C^i_\mu C^j_\nu[X^i,X^j]^m.
\ee

Taking a super-variation on the equations of motion for the fermionic fields (\ref{EOMp3}), one obtains the equations of motion for the scalar fields
\e
0&=&D^2X^i_m+ 2C^j_\mu[D^\mu X^i,X^j]_m-C^j_\mu[D^\mu X^j,X^i]_m+C^{\mu j}C^k_\mu[[X^i,X^k],X^j]_m.
\ee

In summary, we have the equations of motion:
\e\nonumber
0&=&D^2X^i_m+ 2C^j_\mu[D^\mu X^i,X^j]_m-C^j_\mu[D^\mu X^j,X^i]_m+C^{\mu j}C^k_\mu[[X^i,X^k],X^j]_m,\nonumber\\
0&=&\Gamma^\mu D_\mu\psi_{m-}+\Gamma^\mu C^i_\mu[\psi_{-},X^i]_m,\label{alleom}\\
0&=&F^m_{\mu\nu}+2C^i_{[\mu}D_{\nu]}(X^i)^m+C^i_\mu C^j_\nu[X^i,X^j]^m,\nonumber\\
0&=&\partial_\nu C^{\mu i}.\nonumber
\ee
%\subsection{Summary of the Theories}
The set of equations (\ref{alleom}) transform into themselves under the supersymmetry transformations (\ref{susy}). Eqs. (\ref{susy}) and (\ref{alleom}) are main results of this work. The superconformal equations of
motion for a $(2,0)$ theory can be developed via twistor techniques \cite{Saemann2, Saemann3}. It would be very interesting to construct the equations of motion of this paper using a twistor approach.

If we make the replacements
\e
\psi_-\rightarrow\psi_+,\quad \ep_+\rightarrow\ep_-
\ee
in (\ref{susy}) and (\ref{alleom}), where $\psi_+=\frac{1}{2}(1+\Gamma_{6789})\psi$ and $\ep_-=\frac{1}{2}(1-\Gamma_{6789})\ep$,
we will obtain a \emph{new} class of $(1,0)$ hypermultiplet theories.

\section{Acknowledgement}
This work is supported in part by the National Science Foundation
of China (NSFC) under Grant No. 11475016, and supported partially by the Ren-Cai Foundation of Beijing Jiaotong University through Grant No. 2013RC029, and supported partially by the Scientific Research Foundation for Returned Scholars, Ministry of Education of China.

\appendix

\end{document}